\documentclass[fleqn,usenatbib]{mnras}

\usepackage{newtxtext,newtxmath}
\usepackage[T1]{fontenc}
\usepackage{graphicx}	% Including figure files
\usepackage{amsmath}	% Advanced maths commands

\newcommand*{\dyn}{\mbox{{\mdseries \textit{BHNS-Dyn}}}}  
\newcommand*{\dynwind}{\mbox{{\mdseries \textit{BHNS-DynWind}}}}

\newcommand{\revised}[1]{#1}
\newcommand{\review}[1]{#1}%{{\textbf{{#1}}}}

\usepackage{xcolor}

\title[]{Polarized kilonovae from black hole-neutron star mergers}

\author[Bulla et al.]{M. Bulla,$^{1}$\thanks{Email: mattia.bulla@fysik.su.se} K. Kyutoku,$^{2,3,4}$ M. Tanaka,$^{5}$ S. Covino,$^{6}$ J. R. Bruten,$^7$ T. Matsumoto,$^{8,9,10}$ \newauthor   J. R. Maund,$^7$ V. Testa$^{11}$ and K. Wiersema$^{12,13}$
\\
$^1$Nordita, KTH Royal Institute of Technology and Stockholm University, Roslagstullsbacken 23, SE-106 91 Stockholm, Sweden\\
$^2$Department of Physics, Kyoto University, Kyoto 606-8502, Japan\\
$^3$Center for Gravitational Physics, Yukawa Institute for Theoretical Physics, Kyoto University, Kyoto 606-8502, Japan\\
$^4$Interdisciplinary Theoretical and Mathematical Sciences Program (iTHEMS), RIKEN, Wako, Saitama 351-0198, Japan\\
$^5$Astronomical Institute, Tohoku University, Sendai 980-8578, Japan \\
$^6$Istituto Nazionale di Astrofisica / Brera Astronomical Observatory, via Bianchi 46, 23807 Merate (LC), Italy\\
$^7$Department of Physics and Astronomy University of Sheffield Hicks Building, Hounsfield Road Sheffield S3 7RH, UK\\
$^8$Racah Institute of Physics, Hebrew University, Jerusalem 91904, Israel\\
$^9$Research Center for the Early Universe, Graduate School of Science, University of Tokyo, Tokyo 113-0033, Japan\\
$^{10}$Department of Physics, Graduate School of Science, University of Tokyo, Tokyo 113-0033, Japan\\
$^{11}$Istituto Nazionale di Astrofisica/Osservatorio Astronomico di Roma, Monte Porzio
Catone, Italy\\
$^{12}$Department of Physics, University of Warwick, Coventry CV4 7AL, United Kingdom\\
$^{13}$School of Physics and Astronomy, University of Leicester, University Road, Leicester LE1 7RH, United Kingdom
}

\date{Accepted XXX. Received YYY; in original form ZZZ}

\pubyear{2020}

\begin{document}
\label{firstpage}
\pagerange{\pageref{firstpage}--\pageref{lastpage}}
\maketitle 

\begin{abstract}
We predict \revised{linear polarization} for a radioactively-powered kilonova following the merger of a black hole and a neutron star. Specifically, we perform 3-D Monte Carlo radiative transfer simulations for two different models, both featuring a lanthanide-rich dynamical ejecta component from numerical-relativity simulations while only one including an additional lanthanide-free disk wind component. We calculate polarization spectra for nine different orientations at 1.5, 2.5 and 3.5\,d after the merger and in the $0.1-2\,\mu$m wavelength range. We find that both models are polarized at a detectable level 1.5\,d after the merger while show negligible levels thereafter. The polarization spectra of the two models are significantly different. The model lacking a disk wind shows no polarization in the optical, while a signal increasing at longer wavelengths and reaching $\sim1\%-6\%$ at $2\,\mu$m depending on the orientation. The model with a disk-wind component, instead, features a characteristic ``double-peak'' polarization spectrum with one peak in the optical and the other in the infrared. \revised{Polarimetric observations of future events will shed light on the debated neutron richness of the disk-wind component.} The detection of optical polarization would unambiguously reveal the presence of a lanthanide-free disk-wind component, while polarization increasing from zero in the optical to a peak in the infrared would suggest a lanthanide-rich composition for the whole ejecta. Future polarimetric campaigns should prioritize observations in the first $\sim48$\,hours and in the $0.5-2\,\mu$m range, where polarization is strongest, but also explore shorter wavelengths/later times where no signal is expected from the kilonova and the interstellar polarization can be safely estimated.

%Polarimetric observations of BHNS mergers will allow us to distinguish between the two models and 

\end{abstract}

\section{Introduction}
\label{sec:introduction}

Compact object mergers involving at least one neutron star (NS) were long regarded as the most promising scenario for a simultaneous detection of gravitational waves (GWs) and light. This expectation was spectacularly confirmed on August 17, 2017, when the short gamma-ray burst GRB\,170817A \citep{Goldstein2017,Savchenko17} and the radioactively-powered ``macronova'' or ``kilonova'' (hereafter KN) AT2017gfo \citep{Coulter2017} were detected in coincidence with the GW event GW170817 \citep{Abbott2017a}. The worldwide campaign that followed mapped the entire electromagnetic spectrum \citep[e.g.][]{Alexander2017,Andreoni2017,Arcavi2017,Covino2017,Cowperthwaite2017,Evans2017,Drout2017,Haggard2017,Hallinan2017,Kasliwal2017,Margutti2017,Pian2017,Smartt2017,SoaresSantos2017,Tanvir2017,Troja2017,Utsumi2017,Valenti2017} and led to the consensus that a binary neutron star (BNS) merger was at the origin of these signals. The detection of GRB\,170817A provided the smoking gun for the long-thought association between short GRBs and BNS mergers, while the discovery of AT\,2017gfo, a KN powered by the radioactive decay of $r-$process elements synthesized during the coalescence \citep{Li1998}, confirmed that BNS mergers are the prime sites for the production of heavy elements in the Universe \citep[e.g.][]{Kasliwal2019a,Watson2019}.

Following a 19 months break, the Advanced LIGO \citep{Ligo2015} and Advanced Virgo \citep{Acernese2015} interferometers began their third observing run in April, 2019. During this run, the LIGO/Virgo Collaboration (LVC) released 14 real-time public alerts thought to involve at least one NS: 6 BNS mergers and 8 black-hole (BH) - NS mergers. These alerts were followed up by several teams \citep[e.g.][]{Gomez2019,Lundquist2019,Ackley2020,Antier2020,Coughlin2020a,Coughlin2020b,Gompertz2020,Vieira2020,Watson2020b} with the hope to detect an electromagnetic counterpart. The larger distances and poorer localizations of these GW alerts, compared to the golden case of GW170817, required commendable efforts by the various teams to optimize each follow up. However, no convincing electromagnetic counterpart has been reported at the time of writing. 

KN predictions span a wide spectrum depending on the specific properties of the ejected material such as mass, velocity and composition. These, in turn, depend on system parameters like the binary mass ratio, BH spin, NS compactness and equation of state. While the detection of AT\,2017gfo placed important constraints on the physics of BNS mergers (see \citealt{Metzger2019} and \citealt{Nakar2019} for recent reviews), the parameter space of BHNS systems is still largely unexplored observationally (but see \citealt{Yang2015} and \citealt{Jin2015} for a BHNS-KN interpretation of the gamma-ray burst GRB\,060614). In particular, predictions from hydrodynamical merger simulations range from systems where the NS is swallowed whole by the BH to systems where material is ejected both dynamically and from a disk \citep[e.g.][]{Rosswog2005,Foucart2013,Kyutoku2013,Just2015,Kiuchi2015,Kyutoku2015,Fernandez2020}, with the former producing no electromagnetic emission and the latter a luminous KN \citep[e.g.][]{Barbieri2019,Kawaguchi2020,Zhu2020}. While the material ejected dynamically carries the extremely-low electron fraction of the original NS and is thus rich in lanthanides \citep{Korobkin2012}, the exact composition of the material ejected from the post-merger disk is still debated \citep{Siegel2018,Christie2019,Fernandez2019,Miller2019,Fernandez2020,Fujibayashi2020}.

A potential probe to answer some of these open questions is spectropolarimetry. Thomson scattering by free electrons can be an important source of opacity in KNe at early times \citep{Tanaka2018,Banerjee2020} and linearly polarize the escaping radiation. The polarization signal received by a distant observer is sensitive to the system inclination, the geometry and composition of the ejected material, as well as the interplay of the different ejecta components. Assuming an idealized two-component model for the material ejected in BNS mergers, \cite{Bulla2019a} showed that the presence of two distinct ejecta components with different geometries and compositions leads to an overall polarization signal. Specifically, they predicted a maximum polarization level of $\sim0.8\%$ at $\sim7000$\,\AA{} and $1.5$\,d after the merger under favourable (edge-on) inclinations of the system. The polarization signal was found to be smaller for a more face-on orientation \citep[consistent with AT\,2017gfo, e.g.][]{Covino2017,Lamb2019,Troja2019,Hajela2019,Hotokezaka2019}, at different wavelengths and later times. Focusing on the electron-scattering dominated phase in the first few hours following the merger, \cite{Matsumoto2018} and \cite{Li2019b} predicted a polarization signal up to $\sim3\%$ for BNS and BHNS mergers, respectively.

\begin{figure*}
\centering
\includegraphics[width=0.9\textwidth]{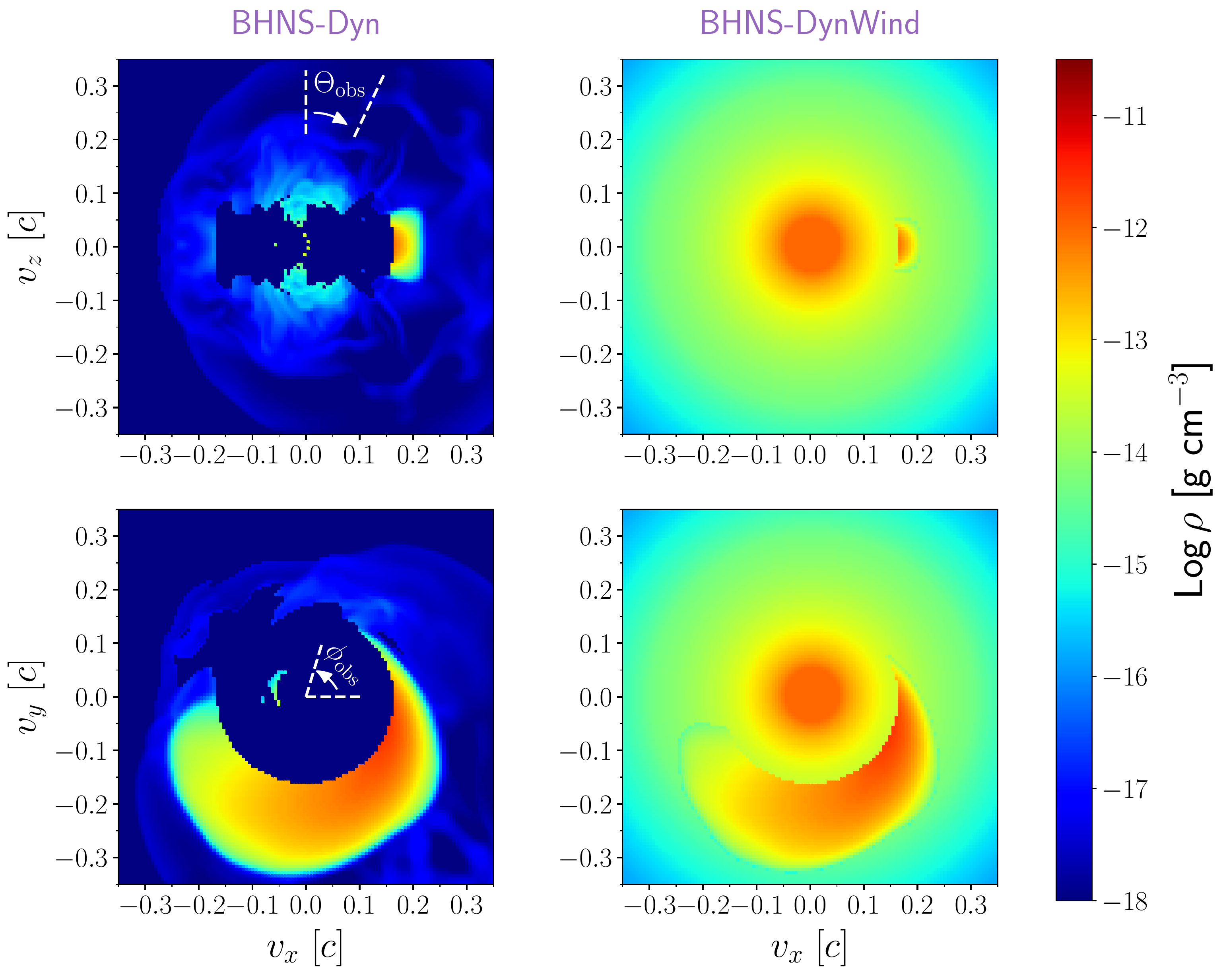}
\caption{Density distribution at 1.5~d after the merger for the \dyn~(left) and \dynwind~(right) models. Top panels show slices in the $x$-$z$ plane while bottom panels slices in the $x$-$y$ (equatorial) plane. \revised{Left panels include the definitions of the viewing angles $\Theta_{\rm obs}$ (top) and $\phi_{\rm obs}$ (bottom).} Low-density regions ($\rho<10^{-18}$\,g\,cm$^{-3}$) close to the center of the \dyn~model are cavities where no unbound material is present in the numerical-relativity simulations of \citet{Kyutoku2015}. \review{All the density maps are shown in velocity space since, under homologous expansion, velocities of each ejecta element do not change with time (i.e. ejecta are freely expanding) and spatial coordinates can be estimated at any epoch $t$ as $(x,y,z)=(v_x,v_y,v_z)\,t$}.}
\label{fig:grid}
\end{figure*}

Here, we follow-up on the work by \cite{Bulla2019a} and carry out Monte Carlo radiative transfer calculations to predict polarization signatures of BHNS KNe at 1.5, 2.5 and 3.5\,d after the merger. \revised{Polarized components coming from other emission mechanisms that were invoked in the case of GW170817 (e.g. cocoon, cocoon breakout) are not included in this analysis.} We employ a numerical-relativity simulation of BHNS mergers from \citet{Kyutoku2015} and also explore the impact of a spherical post-merger disk wind on the predicted signal. We introduce our BHNS models in Section~\ref{sec:models}, while describe the setup of our radiative tranfer calculations in Section~\ref{sec:radtransf}. We summarize our results in Section~\ref{sec:results} and discuss them in Section~\ref{sec:disc}. 

\section{Models}
\label{sec:models}

In this work, we study two separate BHNS models with different ejecta geometries and compositions. The first model, referred to as \dyn, employs 3-D dynamical ejecta from the numerical-relativity simulations of \cite{Kyutoku2015}. Specifically, we adopt the H4-Q3a75 model, which assumes a $4.05\,M_\odot$ BH with the dimensionless spin parameter equal to 0.75, a $1.35\,M_\odot$ NS and the H4 equation of state (\citealt{Lackey2006}, see table II of \citealt{Kyutoku2015} for more details). For our radiative transfer simulations (see Section~\ref{sec:radtransf}), we extrapolate the profile of the dynamical ejecta from $10$\,ms \citep{Kyutoku2015} to $1.5-3.5$\,d after the merger by assuming homologous expansion. This procedure overestimates the expansion of the inner part of the ejecta, which is significantly affected by the gravitational potential of the remnant BH at $10$\,ms after the onset of merger. Although this simplification is crude, we find that it does not affect our results strongly since the high degree of polarization is observed for directions not affected by this assumption (see ${\bf n}_{\#{\rm d}}$ and ${\bf n}_{\#{\rm a}}$ in Section~\ref{sec:results}). Modeling the long-term evolution of the ejecta \citep{Rosswog2014} is necessary for more accurate predictions of the KN and is left for future work. After restricting to the amount of unbound material in the model, we obtain an ejecta mass of $M_{\rm BHNS-Dyn}=M_{\rm dyn}=0.043\,M_\odot$. As shown in the left panels of Fig.~\ref{fig:grid}, the dynamical ejecta are concentrated about the $x$-$y$ (orbital) plane in a crescent-like shape. A lanthanide-rich composition is assumed for this component (see Section~\ref{sec:introduction}), with specific opacities selected from \citet[][see Section~\ref{sec:radtransf}]{Tanaka2018}. The one-component \dyn~model may approximate the emission from massive BHNS binary mergers, for which the dynamical ejecta could dominate the disk wind \citep[][see their fig.~11]{Kyutoku2015}. \revised{Radiative transfer simulations assuming homologous expansion were carried out by \citet{Tanaka2014} for a similar BHNS model without a disk-wind component, although the focus was on predicting light curves and colors and no polarization predictions were made.}

%\kk{I recall that we could add a sentence like "This setup can also be directly compared to simulations of Tanaka et al. 2014." https://ui.adsabs.harvard.edu/abs/2014ApJ...780...31T/abstract although the simulations here are performed with old line lists}

The second model, referred to as \dynwind~and shown in the right panels of Fig.~\ref{fig:grid}, is a variant of the first model where a post-merger disk wind is added to the dynamical ejecta from \cite{Kyutoku2015}. The wind is assumed to be spherical and lanthanide-free in composition (see Section~\ref{sec:radtransf}). The disk mass predicted for the H4-Q3a75 model of \cite{Kyutoku2015} is $M_{\rm disk}=0.3\,M_\odot$, $20\%$--$40\%$ of which is predicted to be \revised{ejected} in the form of a disk wind \citep{Just2015,Siegel2018,Miller2019,Fernandez2019,Fernandez2020,Fujibayashi2020}. Here, we assume that the ejected disk wind is $30\%$ of the disk mass, i.e. $M_{\rm wind}=0.3\,M_{\rm disk}=0.09\,M_\odot$. This material is distributed assuming a broken power-law for the density profile, with $\rho\propto v^{-3}$ for $0.05<v<0.3$c and $\rho\propto v^{-7}$ for $v>0.3$c, where c is the speed of light. This particular density profile is chosen to mimic the one in \cite{Bulla2019a}. A density floor is assumed for the inner regions of the model ($v<0.05$c), adding an extra $0.016\,M_\odot$ of material to the wind ejecta. Combining the wind and the dynamical component, the total ejecta mass of the \dynwind~model is equal to $M_{\rm BHNS-DynWind}=0.149\,M_\odot$. When combining the dynamical ejecta and disk wind, we assume each grid cell to be dominated by the denser component. \revised{This approximation reduces the mass of the dynamical ejecta in the \dynwind~model only by $1.35 \times 10^{-4} M_\odot$ ($\sim0.3\%$) compared to that in the \dyn~model.}

\section{Radiative transfer simulations}
\label{sec:radtransf}

\revised{Linear polarization is} calculated for the \dyn~and \dynwind~models using the 3-D Monte Carlo radiative transfer code \textsc{possis} \citep{Bulla2019b}\footnote{\review{Validation tests of \textsc{possis} can be found at \url{https://github.com/mbulla/kilonova_models/tree/master/pol_tests}, including comparisons to analytic and numerical solutions in optically-thin and optically-thick atmospheres \cite[see also][]{Bulla2015}.}}. Briefly, \textsc{possis} discretises the radiation in $N_{\rm ph}$ Monte Carlo quanta and follows them as they diffuse out the ejecta and interact with matter. \revised{The linear polarization is described in terms of the Stokes vector ${\bf S}=(I,Q,U)$\footnote{\revised{Here we neglect circular polarization and set the Stokes parameter $V=0$. As shown by \cite{Chandrasekhar1960}, circular polarization satisfies a transfer equation that is independent from linear polarization in the absence of magnetic field.}} or the normalized dimensionless Stokes vector ${\bf s}={\bf S}/I = (1,q,u)$}. For each Monte Carlo quantum, the normalized Stokes vector is initialized to ${\bf s}_0=(1,0,0)$ and updated after each interaction with matter. Specifically, electron scattering is assumed to polarize while bound-bound transitions to depolarize the radiation (see \citealt{Bulla2015} an d \citealt{Bulla2019b} for more details). To speed up the calculations, synthetic observables including polarization are extracted using the Event Based Technique (EBT) described in \cite{Bulla2015}.

In this work, we carry out radiative transfer simulations using $N_{\rm ph}=10^7$ Monte Carlo quanta. Given that the models are symmetric about the $x-y$ equatorial plane (see top panels of Fig.~\ref{fig:grid}), we restrict our investigation to the northern hemisphere ($z\geq0$) and select nine observer orientations \revised{defined by their angles $\Theta_{\rm obs}$ and $\phi_{\rm obs}$ (see left panels of Fig.~\ref{fig:grid} for a definition of these angles).} 
%as shown in the left panels of Fig.~\ref{fig:dyn} (\dyn) and Fig.~\ref{fig:dynwind} (\dynwind). 
The first viewing angle, ${\bf n}_{1{\rm a}}$, is chosen along the $z$ axis ($\cos\Theta_{\rm obs}=1$ and $\phi_{\rm obs}=0^\circ$). Four viewing angles are then selected $60^\circ$ away from the pole ($\cos\Theta_{\rm obs}=0.5$) and with different azimuthal angles: ${\bf n}_{2{\rm a}}$ ($\phi_{\rm obs}=0^\circ$), ${\bf n}_{2{\rm b}}$ ($\phi_{\rm obs}=90^\circ$), ${\bf n}_{2{\rm c}}$ ($\phi_{\rm obs}=180^\circ$) and ${\bf n}_{2{\rm d}}$ ($\phi_{\rm obs}=270^\circ$). Similarly, four viewing angles are selected in the equatorial plane ($\cos\Theta_{\rm obs}=0$): ${\bf n}_{3{\rm a}}$ ($\phi_{\rm obs}=0^\circ$), ${\bf n}_{3{\rm b}}$ ($\phi_{\rm obs}=90^\circ$), ${\bf n}_{3{\rm c}}$ ($\phi_{\rm obs}=180^\circ$) and ${\bf n}_{3{\rm d}}$ ($\phi_{\rm obs}=270^\circ$). The angles $\phi_{\rm obs}$ and $\Theta_{\rm obs}$ are chosen \revised{to sample uniformly the $2\pi$ solid angle}. \revised{We note that the dynamical ejecta are distributed in a crescent-like shape centered at $\phi_{\rm obs}\sim270^\circ$ (see bottom left panel of Fig.~\ref{fig:grid}).}

\begin{figure*}
\centering
\includegraphics[width=0.95\textwidth]{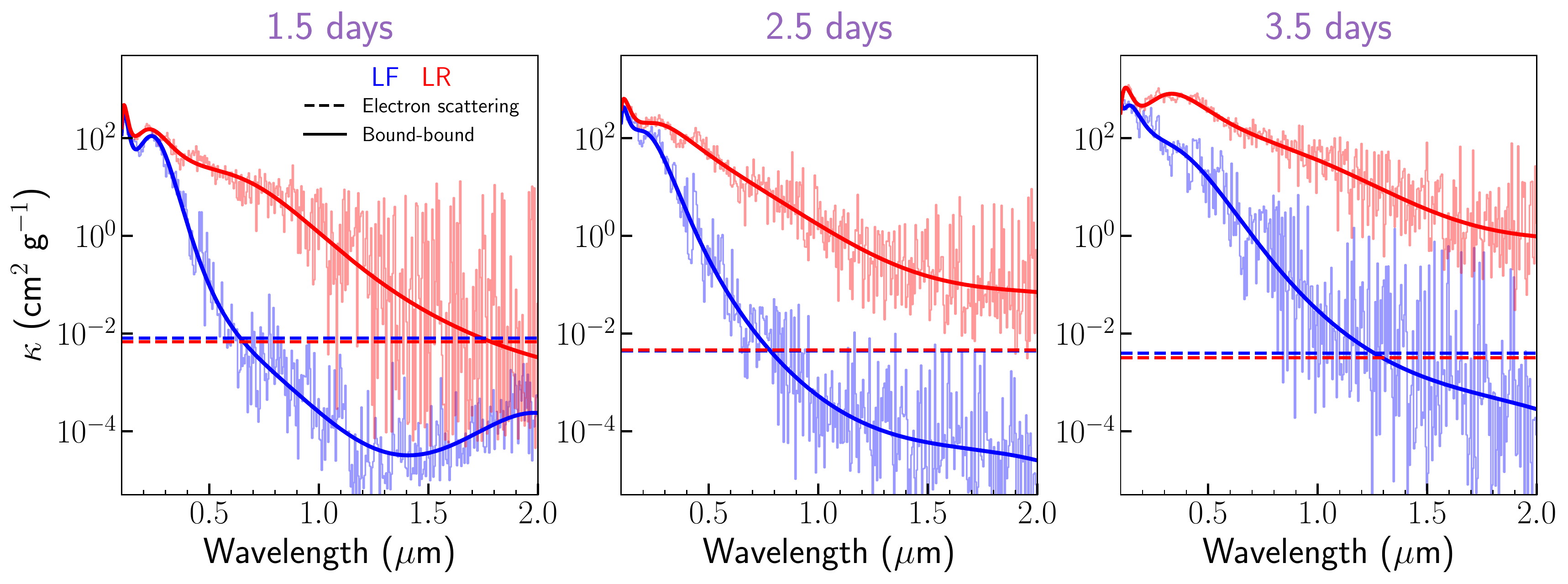}
\caption{Lanthanide-free (blue) and lanthanide-rich (red) opacities from \citet{Tanaka2018} adopted in this work. Electron scattering opacities (horizontal dashed lines) and a 10th-order polynomial fit (thick solid line) to the bound-bound opacities (thin solid line) are implemented in \textsc{possis} and used in the simulations. Opacities at the three different epochs (1.5, 2.5 and 3.5\,d since merger) are shown from left to right. \review{The steep decrease in bound-bound opacities with wavelength is due to the number of transitions being larger at bluer wavelengths.}  }
\label{fig:opac}
\end{figure*}

Polarization levels are computed at three individual epochs: 1.5, 2.5 and 3.5~d since merger. \revised{These epochs correspond to the first three polarimetric observations of AT\,2017gfo \citep{Covino2017} and are chosen for ease of comparison with the results in \cite{Bulla2019a}.} Coordinates and densities for the \dyn~and \dynwind~models are remapped to the desired epochs according to homologous expansion (i.e. assuming constant velocity $v$ and density dropping as $\rho\propto t^{-3}$). 

Two main changes are made compared to the calculations presented in \cite{Bulla2019a} for BNS mergers. First, the initial location of the Monte Carlo quanta is sampled from the ejecta mass distribution (i.e. more quanta are created at higher compared to lower densities). \revised{This approach is a better description of the underlying physics, where a fraction of the $\gamma$-ray photons from the radioactive decay of $r-$process elements are absorbed by matter and re-emitted at longer wavelengths as a thermal component \citep{Li1998}.} This assumption improves on the approximation adopted in \citet{Bulla2019a}, where Monte Carlo quanta were emitted only from a spherical photosphere pre-defined at some given velocity/radius. Second, the full wavelength information of opacities is used to predict polarization spectra between 0.1 and $2\,\mu$m rather than polarization levels at individual wavelengths. In particular, \revised{realistic} opacities \revised{based on atomic calculations} from \cite{Tanaka2018} are used for both lanthanide-free (lf) and lanthanide-rich (lr) compositions at each epoch.  \revised{At the wavelengths investigated in this study, electron scattering and bound-bound interactions are the main source of opacity.} Electron-scattering and bound-bound opacities used in this work are shown in Fig.~\ref{fig:opac}. \revised{Their relative contribution is sensitive to the density/temperature conditions within the ejecta, with electron scattering (bound-bound) opacities typically decreasing (increasing) with time as the ejecta expand, cool down, and the different elements recombine to lower-ionization stages.} \review{The steep decrease in bound-bound opacities with wavelength is due to the number of transitions being larger at bluer wavelengths.} The values adopted for the electron scattering opacities at the different epochs are $(\kappa^{\rm lf}_{\rm es},\kappa^{\rm lr}_{\rm es})_{1.5\,{\rm d}}=(0.0080,0.0068)\,{\rm cm}^2\,{\rm g^{-1}}$, $(\kappa^{\rm lf}_{\rm es},\kappa^{\rm lr}_{\rm es})_{2.5\,{\rm d}}=(0.0044,0.0046)\,{\rm cm}^2\,{\rm g^{-1}}$ and $(\kappa^{\rm lf}_{\rm es},\kappa^{\rm lr}_{\rm es})_{3.5\,{\rm d}}=(0.0040,0.0032)\,{\rm cm}^2\,{\rm g^{-1}}$. 
%\revised{Electron scattering opacities decrease with time following the decrease in the ionization degree of the ejecta \citep{Tanaka2018,Banerjee2020}.}
Bound-bound opacities $(\kappa^{\rm lf}_{\rm bb},\kappa^{\rm lr}_{\rm bb})$ at each wavelength and time are computed through 10th-order polynomial fits of the opacities from \citet[][see thick solid lines in Fig.~\ref{fig:opac}]{Tanaka2018}.

\section{Results}
\label{sec:results}

\begin{figure*}
\centering
\includegraphics[width=0.965\textwidth]{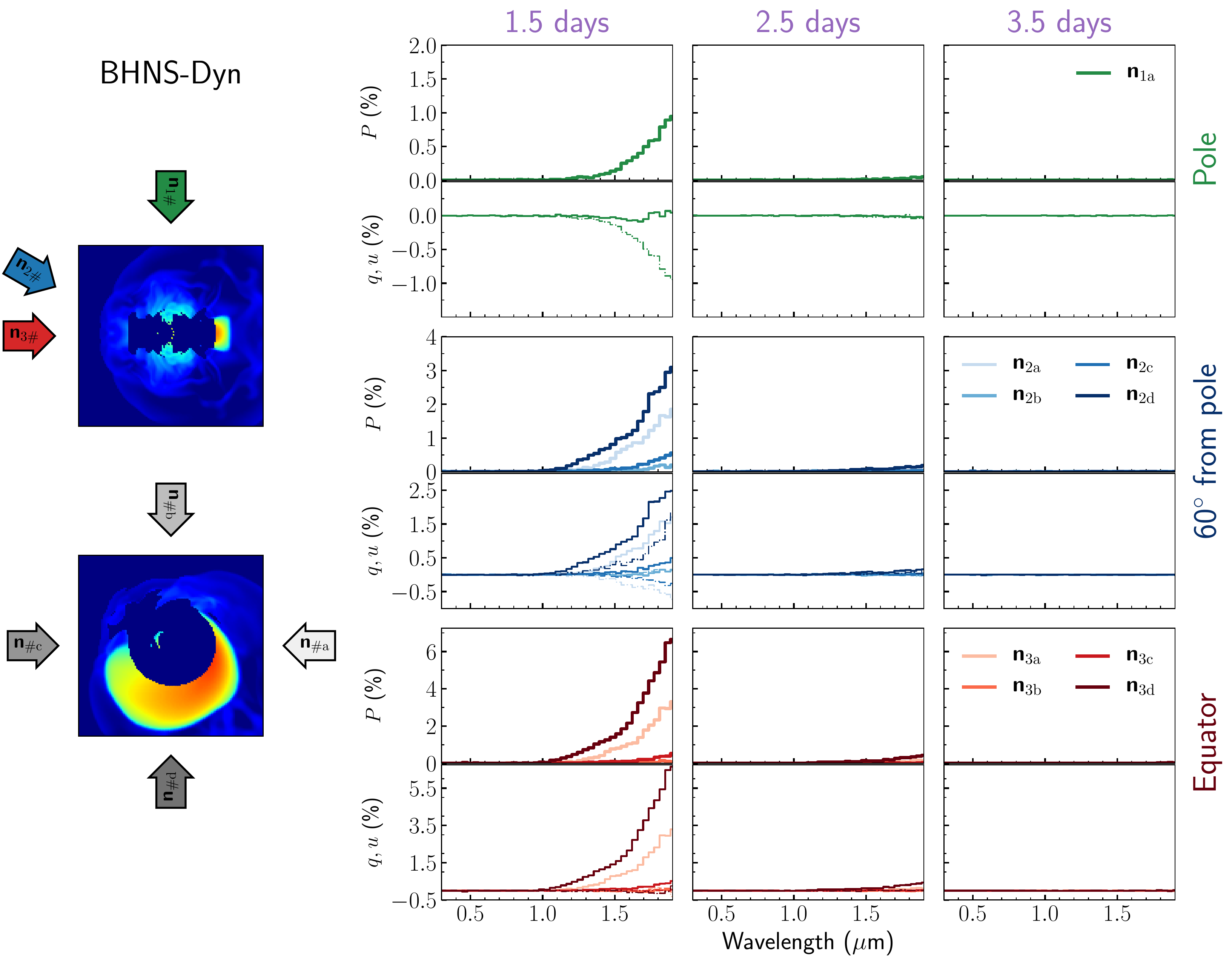}
\caption{Polarization spectra for the BHNS-Dyn model. \textit{Left panels}: location of the 9 viewing angles with respect to the ejecta geometry as viewed in the $x$-$z$ (top) and $x$-$y$ (bottom) planes. \textit{Lower panels}: $q$ (solid thin lines), $u$ (dashed thin lines) and $P$ (thick lines) spectra for all the different viewing angles. Spectra are grouped in different rows according to the observer polar angle $\Theta_{\rm obs}$: the top row refers to $\cos\Theta_{\rm obs}=1$ (${\bf n}_{1{\rm a}}$), the middle row to $\cos\Theta_{\rm obs}=0.5$ (${\bf n}_{2{\rm a,b,c,d}}$) and the bottom row to $\cos\Theta_{\rm obs}=0$ (${\bf n}_{3{\rm a,b,c,d}}$). Spectra at different epochs are shown in different columns (1.5, 2.5 and 3.5~d since merger from left to right). Spectra have been re-binned of  a factor of two for presentation purposes. Note that the y-axis scale is different in different rows. }
\label{fig:dyn}
\end{figure*}

\begin{figure*}
\centering
\includegraphics[width=0.965\textwidth]{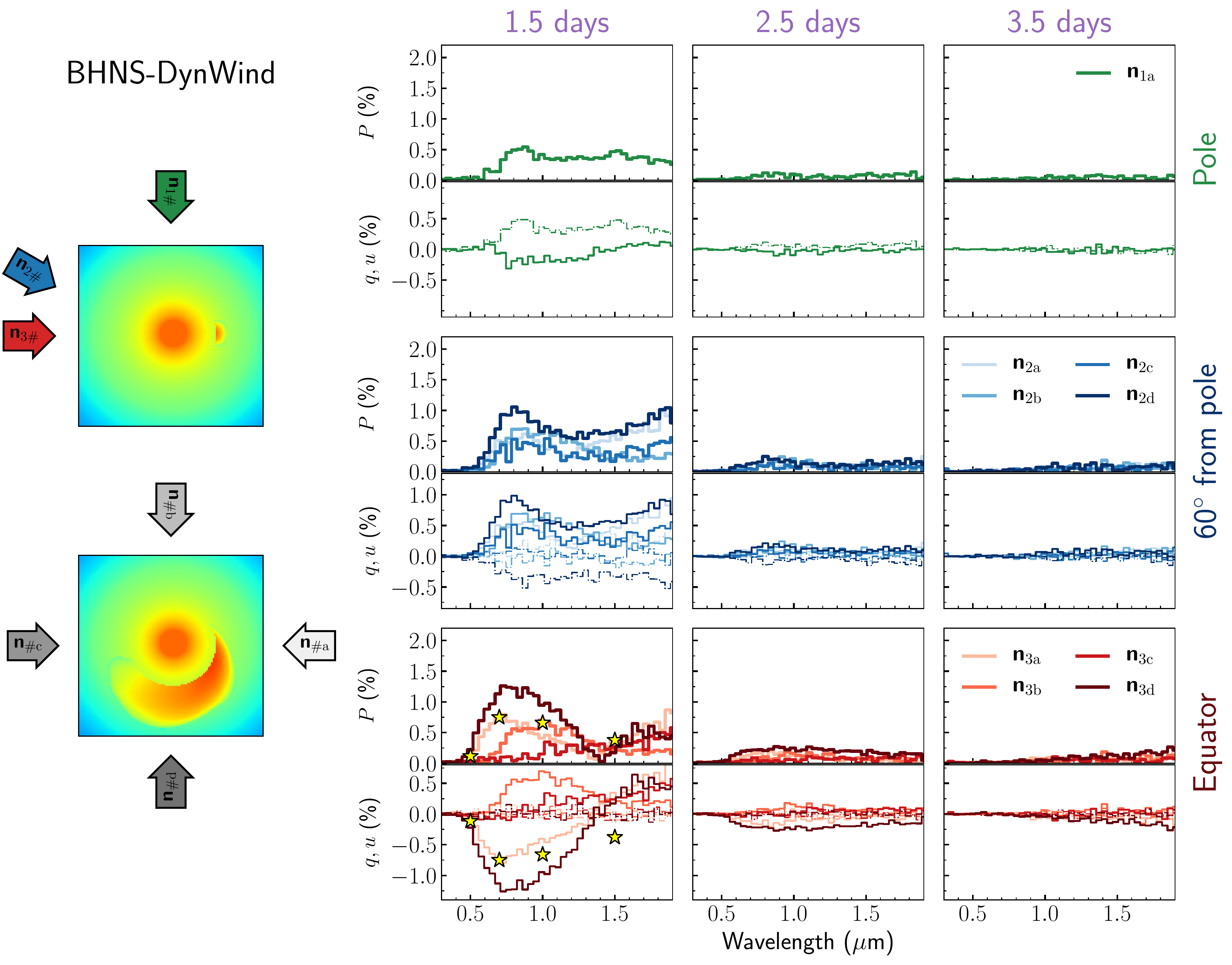}
\caption{Same as Fig.~\ref{fig:dyn} but for the BHNS-DynWind model. The yellow stars in the bottom left panel mark the $P$ (top) and $q$ (bottom) values obtained by \citet{Bulla2019a} for a BNS merger.}
\label{fig:dynwind}
\end{figure*}

In the following, we summarize the results of the simulations described in Section~\ref{sec:radtransf}. We begin with the \dyn~model in Section~\ref{sec:dynamical} and then discuss the \dynwind~model in Section~\ref{sec:dynwind}. The implications of our results in terms of observations will be discussed in Section~\ref{sec:disc}.

\subsection{Dynamical ejecta}
\label{sec:dynamical}

Polarization spectra computed for the \dyn~model are shown in Fig.~\ref{fig:dyn}. Viewing angles are grouped in different rows depending on the polar angle $\Theta_{\rm obs}$ (polar to equator from top to bottom), while different columns show predictions at different epochs (1.5, 2.5 and 3.5~d from left to right). The polarization signal shows a clear wavelength-, angular- and time-dependence. We address each of these three behaviours below.

We start by focusing on the polar orientation ($\cos\Theta_{\rm obs}=1$, ${\bf n}_{1{\rm a}}$) at 1.5~d after the merger, shown in the top-left panel of Fig.~\ref{fig:dyn}. The predicted level is strongly wavelength dependent, with polarization consistent with zero below $\sim1.2\,\mu$m and increasing moving to longer wavelengths. The polarization level reaches $P\sim1\%$ at $\sim2\,\mu$m, at the boundary of the simulated wavelength domain. The wavelength-dependence of the signal is a direct consequence of the adopted opacities and it is controlled by the ratio $\kappa_{\rm es}/\kappa_{\rm bb}$ between polarizing electron-scattering contributions and depolarizing bound-bound interactions (see Section~\ref{sec:radtransf}). In the \dyn~model, the ejecta are assumed to be lanthanide-rich and thus characterized by relatively high bound-bound opacities. As shown in the left panel of Fig.~\ref{fig:opac} (see red lines), $\kappa_{\rm es}/\kappa_{\rm bb}\ll1$ below $\sim1.2\,\mu$m, hence the negligible polarization levels obtained in this wavelength range. Starting from $\sim1.2\,\mu$m, the $\kappa_{\rm es}/\kappa_{\rm bb}$ ratio reaches values larger than 0.1 and eventually exceed unity, hence the rise in polarization moving farther into the infrared.

The signal is strongly viewing-angle dependent due to  the large-scale asymmetries in the \dyn~model and the resulting projection effects. Polarization levels are found to be the highest for orientations facing the crescent-like ejecta distribution, namely $\phi_{\rm obs}=0^\circ$ (${\bf n}_{2{\rm a}}$ and ${\bf n}_{3{\rm a}}$) and $\phi_{\rm obs}=270^\circ$ (${\bf n}_{2{\rm d}}$ and ${\bf n}_{3{\rm d}}$), cf. with bottom-left panel of Fig.~\ref{fig:dyn}. \revised{As mentioned in Section~\ref{sec:models}, these viewing angles are the least affected by the assumption of homologous expansion in our calculations.} At $\phi_{\rm obs}=0^\circ$, the polarization level at $\sim2\,\mu$m increases from $\sim1\%$ at the Pole (${\bf n}_{1{\rm a}}$) to $\sim1.5\%$ at $\Theta_{\rm obs}=60^\circ$ (${\bf n}_{2{\rm a}}$) and $\sim3\%$ at the Equator (${\bf n}_{3{\rm a}}$). At $\phi_{\rm obs}=270^\circ$, the polarization peak at $\sim2\,\mu$m reaches even higher levels: $\sim3\%$ at $\Theta_{\rm obs}=60^\circ$ (${\bf n}_{2{\rm d}}$) and $\sim6\%$ at the Equator (${\bf n}_{3{\rm d}}$). In contrast, the polarization is much lower ($\lesssim0.5\%$) for orientations at the other side of the crescent-like ejecta distribution ($\phi_{\rm obs}=90^\circ$ and $\phi_{\rm obs}=180^\circ$) as they receive fewer polarizing contributions. \revised{However, the results for these directions should be taken with care since the assumption of the homologous expansion with a central cavity may introduce significant artificial effects (see Section~\ref{sec:models})}. \review{Quantifying the impact of this assumption on the polarization predicted along these orientations is left for future studies. Here we note that lanthanide-rich material filling the cavity would be associated with high densities/opacities, with radiation created in these regions diffusing out on long timescales and affecting only marginally the early-time polarization predicted in this study.}

The time-dependence of the polarization signal is very strong. At 2.5~d after the merger, i.e. only 1~d after the first simulated epoch, the levels are negligible at all wavelengths and for all the different viewing angles ($\lesssim0.3\%$ at $2\,\mu$m). Polarization signals are even weaker and essentially consistent with zero at 3.5~d after the merger. The extremely rapid drop in polarization is a consequence of the time evolution of the $\kappa_{\rm es}/\kappa_{\rm bb}$ ratio and of the electron-scattering optical depth $\tau_{\rm es}$. As shown in Fig.~\ref{fig:opac}, the bound-bound opacities rapidly increase with time as the ejecta cool and atoms recombine, while the variation in electron scattering opacity is much smaller. Therefore, the $\kappa_{\rm es}/\kappa_{\rm bb}$ ratio decreases extremely fast, with $\kappa_{\rm es}/\kappa_{\rm bb}\sim0.05$ ($0.003$) at $\sim2\,\mu$m and 2.5 (3.5)\,d after the merger. In addition, the density drop due to the expansion (a factor of $\sim5$ between the first and the second epoch and of $\sim13$ between the first and the third epoch, $\rho\propto t^{-3}$) causes the electron-scattering optical depth $\tau_{\rm es}$ to rapidly decrease with time. These effects lead to a rapid decrease in polarizing contribution and therefore resulting polarization signal.

\subsection{Dynamical + wind ejecta}
\label{sec:dynwind}

Polarization spectra computed for the \dynwind~model are shown in Fig.~\ref{fig:dynwind}, with predictions for different viewing angles and epochs organized in different panels as in Fig.~\ref{fig:dyn} (see Section~\ref{sec:dynamical}). Similarly to what is seen in the \dyn~model, the polarization signal predicted for the \dynwind~model depends on wavelength, viewing angle and time. 

Again, we begin by focusing on the polar viewing angle ($\cos\Theta_{\rm obs}=1$, ${\bf n}_{1{\rm a}}$) at 1.5~d after the merger, shown in the top-left panel of Fig.~\ref{fig:dynwind}. The overall polarization level is $\lesssim0.5\%$ at all wavelengths investigated and shows a characteristic ``double-peak'' structure: one peak is seen in the optical ($\sim0.7\,\mu$m) and one in the infrared ($\sim1.5\,\mu$m). This behaviour can be understood by inspecting the wavelength-dependence of opacities from Fig.~\ref{fig:opac} (see left panel). While the \dyn~model is characterized by lanthanide-rich opacities, the \dynwind~model features both lanthanide-free opacities from the wind and lanthanide-rich opacities from the dynamical ejecta. The interplay between the two components is responsible for shaping the polarization spectrum, which can be broadly summarized by looking at three separate wavelength ranges. In the first range, $\lambda\lesssim0.5\,\mu$m, $\kappa_{\rm es}/\kappa_{\rm bb}\ll1$ for both components and thus the polarization signal is consistent with zero. In the second range, $0.5\lesssim\lambda\lesssim1.2\,\mu$m, $\kappa^{\rm lf}_{\rm es}/\kappa^{\rm lf}_{\rm bb}\gtrsim1$ and $\kappa^{\rm lr}_{\rm es}/\kappa^{\rm lr}_{\rm bb}\ll1$ and thus the observer receives polarizing contributions from the spherical wind and depolarizing contributions from the dynamical ejecta. While all the polarizing contributions would cancel to zero if the spherical wind was the only component in the model, the presence of a depolarizing asymmetric component (i.e. the dynamical ejecta) leads to an overall polarization signal. This effect, which is maximized at $\sim0.7\,\mu$m, is the same as the one first discussed by \cite{Bulla2019a} (see also Section~\ref{sec:disc}). In the third range, $\lambda\gtrsim1.2\,\mu$m, $\kappa_{\rm es}/\kappa_{\rm bb}\gtrsim1$ in both components and thus the ejecta are dominated by electron scattering. While this resulted in large polarization signals for the strongly asymmetric \dyn~model, the infrared polarization predicted in the \dynwind~model is modest due to the configuration approaching a pure-electron scattering case in a spherical photosphere \citep{Bulla2019a}.

The viewing-angle dependence of the polarization signal is shown in the different rows of Fig.~\ref{fig:dynwind}. In general, the polarization spectra continue to display a double-peak profile as the one described above. In the optical, the strongest polarization is seen for orientations at $\phi_{\rm obs}=0^\circ$ and $\phi_{\rm obs}=270^\circ$ as it was the case in the \dyn~model. This is because the dynamical ejecta are distributed preferentially towards these orientations, thus blocking and breaking the cancellation of polarizing contribution from the wind more effectively (see above). Quantitatively, a peak of $P\sim1.2\%$ is found at $\sim0.7\,\mu$m when the system is viewed from the ${\bf n}_{2{\rm d}}$ and ${\bf n}_{3{\rm d}}$ orientation. In contrast, the infrared polarization is less viewing-angle dependent reflecting the overall symmetry achieved in the polarizing contributions at these wavelengths (see above).

The \dynwind~model is also characterized by a rapid depolarization of the signal, with levels below $\sim0.3\%$ at 2.5 and 3.5\,d after the merger. As explained in Section~\ref{sec:dynamical}, this behaviour is caused by both the $\kappa_{\rm es}/\kappa_{\rm bb}$ ratio and the (polarizing) electron-scattering opacity decreasing with time. We note, however, that the presence of a lanthanide-free component in the \dynwind~model leads to relatively higher levels compared to those in the \dyn~model, especially \revised{at wavelengths shorter than} $\sim1.3\,\mu$m.

\section{Discussion and conclusions}
\label{sec:disc}

Using the Monte Carlo radiative transfer code \textsc{possis}, we predict polarization spectra of two 3-D BHNS models as viewed from nine different orientations. While both models feature the presence of a lanthanide-rich dynamical ejecta component \citep{Kyutoku2015}, the \dynwind~model includes also a lanthanide-free spherical wind component that is not present in the \dyn~model.

A clear polarization signal is found in the wavelength range investigated ($0.1-2\,\mu$m), reaching $\sim6\%$ for the \dyn~model viewed from an equatorial viewing angle. The signal is restricted to early epochs, $1.5\,$d after the merger, and disappears within a day or two. This indicates that a polarization signal could be detected in future KNe resulting from the merger of a NS and a BH. Our quantitative predictions call for spectropolarimetric observations that are rapid and cover a wavelength range as wide as possible. While prioritizing observations in the first $\sim48$~hours and in the $0.5-2\,\mu$m range, where the signal is expected to be the highest, the best strategy would be to obtain additional observations at later times and shorter wavelengths. This would allow to explore epochs and wavelengths where the intrinsic signal is predicted to be negligible and thus the interstellar polarization can be robustly estimated and removed from all the observations (see also \citealt{Bulla2019a}). \revised{In this respect, we note that the predicted wavelength dependence in the polarization of the KN is quite different from that seen in (Galactic and host-galaxy) interstellar dust, making the separation of these two components easier even for low signal-to-noise data \citep{Patat2015}.}

The polarization spectra are significantly different between the two models. While the \dyn~model is polarized only in the infrared ($\lambda\gtrsim1.2\,\mu$m), the \dynwind~model features  a ``double-peak'' polarization spectrum with a peak in the optical ($\sim0.7\,\mu$m) and another in the infrared ($\sim1.3-2\,\mu$m depending on the viewing angle). \revised{Even in cases where the two peaks are difficult to resolve in data (e.g. because of low amplitudes), their distinction should be facilitated by the corresponding polarization angles (i.e. $q/u$) changing differently with time and wavelength (see e.g. bottom panels of Fig.~\ref{fig:dynwind}).} The clear difference between the two models is ascribed to the presence of a lanthanide-free spherical wind component\revised{, indicating that polarimetric data of future events might constrain the neutron richness of the disk-wind component in BHNS mergers, a property that is yet not settled in the literature \citep{Siegel2018,Christie2019,Fernandez2019,Miller2019,Fernandez2020,Fujibayashi2020}}. Specifically, the detection of a polarization signal at optical wavelengths would be a smoking-gun for the presence of a lanthanide-free disk-wind component, \revised{especially if accompanied by an early-blue emission as in AT\,2017gfo}. Alternatively, a non-detection in the optical together with a strong ($\gtrsim2\%$ at $\sim2\,\mu$m) signal in the infrared would point to a lanthanide-rich composition for the entire ejecta. \revised{Knowledge about the composition can constrain the disk mass and the launching mechanism/timescales of the post-merger ejecta \citep{Fujibayashi2020}.}

Our predictions call for (spectro)polarimetric observations extending as much as possible into the infrared domain. While infrared polarimeters are rare on large telescopes, \cite{Tinyanont2019} showed that high signal-to-noise polarimetry of bright transients should be feasible at these wavelengths with the WIRC+Pol instrument mounted on the 200 inch ($\sim5$m) Hale Telescope at Palomar Observatory. \revised{As shown by Tinyanont et al. (submitted), the WIRC+Pol instrument can observe sources as faint as 14.5 mag in the $J$ band to a polarimetric accuracy of d$P\sim0.1\%$ per spectral channel in less than 2 hours. This accuracy would be sufficient to detect a polarization signal at $\sim1$\,d after the merger in a nearby ($\sim10$\,Mpc) AT2017gfo-like KN. An even more promising option is offered by the IRCS instrument at the $8.2$m Subaru telescope \citep{Watanabe2018}. A nominal\footnote{\url{https://www.naoj.org/Observing/Instruments/IRCS/polarimetry/polarimetry.html}} accuracy of d$P\sim0.5\,\%$ can be achieved with 1-hour exposure for $J\sim20.7$\,mag, $H\sim20.0$\,mag and $K\sim19.5$\,mag, which would enable the detection of high polarization signals as those predicted here in a AT2017gfo-like KN up to $\sim150-200$\,Mpc. Additional near-infrared polarimeters suitable for bright transients include LIRIS (Long-slit Intermediate Resolution Infrared Spectrograph) on the 4.2m William Herschel Telescope \citep[WHT, see e.g.][]{Manchado,Wiersema2012} and SofI on the 3.6m New Technology Telescope \citep[NTT, see e.g.][]{Higgins2019}.} 
%Additional near-infrared polarimeters include Gemini/NIRI \citep{Hodapp2003} and} \textcolor{red}{others? VLT?}.

Within a given model, a clear viewing-angle dependence is found in the polarization signal. This brings the potential to pin-point the system inclination of future BHNS mergers using spectropolarimetric observations, especially when combined with constraints from different probes (e.g. KN optical/near-infrared spectra and light curves, radio observations). \revised{In the future, systematic studies of polarization from various configurations of the ejecta components will be crucial to determine the viewing angle accurately. Estimating the inclination of NS mergers is important to understand the physics of these systems and in particular their ejecta distribution.}

%Constraining the inclination of future BHNS and BNS mergers would be important to e.g. break the distance-inclination degeneracy in the GW signal \citep{Abbott2017b} and reduce the uncertainties in the Hubble constant measurements \citep{Guidorzi2017,Hotokezaka2019,Dhawan2020}.

Our polarization predictions can be compared \review{to those in the literature. \cite{Matsumoto2018} and \cite{Li2019b} predicted polarization arising from a fast ejecta component ($\gtrsim$0.5c) in BNS and BHNS, respectively, where free neutrons survive and their $\beta$-decay electrons dominate the scattering opacity in the first few hours after the merger. This high-velocity component is neglected in the present work and thus a one-to-one comparison with \cite{Matsumoto2018} and \cite{Li2019b} is not possible. A more meaningful comparison can be done with \cite{Bulla2019a}, which focused on BNS mergers and assumed} a two-component (wind+dynamical) ejecta geometry similar to that in the \dynwind~model presented here. However, the dynamical ejecta in their model have a torus-like shape while ours a crescent-like shape covering only $\sim$\,half of the azimuthal angle. The assumed axial symmetry in the \cite{Bulla2019a} model leads to a polarization signal that is zero at the pole, when the ejecta are symmetric in projection, and increases monotonically towards the equator (see their fig. 2). Observationally, this means that a BNS KN would be unpolarized or very lowly polarized when viewed closed to face-on, as GW170817/AT2017gfo was \citep[e.g.][]{Covino2017,Lamb2019,Troja2019,Hajela2019,Hotokezaka2019}. In contrast, our \dynwind~model is not axially symmetric and thus it is polarized in any orientation investigated. This suggests that BHNS systems might be easier to detect in polarization compared to BNS, although we note that this statement depends on whether dynamical ejecta deviating substantially from the axial symmetry can be produced in BNS mergers and/or whether a lanthanide-free component is present in BHNS mergers. Finally, we note that predictions from \cite{Bulla2019a} for an equatorial viewing angle match remarkably well the values obtained by \dynwind~model when viewed from the  ${\bf n}_{3{\rm a}}$ orientation (see yellow stars in Fig.~\ref{fig:dynwind}). This is not too surprising since the torus-like geometry in the \cite{Bulla2019a} model is relatively similar in projection to the crescent-like shape in our \dynwind~model when viewed from ${\bf n}_{3{\rm a}}$.

\section*{Acknowledgements}
\review{The authors are thankful to the anonymous reviewer for helpful comments that improved this paper.} Koutarou Kyutoku is supported by Japanese Society for the Promotion of Science (JSPS) Kakenhi Grant-in-Aid for Scientific Research (No.~JP16H06342, No.~JP17H01131, No.~JP18H05236, No.~JP19K14720, No.~JP20H00158). 

\section*{Data Availability}

The polarization spectra presented in this work are publicly available at \url{https://github.com/mbulla/kilonova_models}.

\bibliographystyle{mnras}

{\footnotesize
\bibliography{astrobulla.bib}}

\providecommand{\noopsort}[1]{}
\begin{thebibliography}{}
\makeatletter
\relax
\def\mn@urlcharsother{\let\do\@makeother \do\$\do\&\do\#\do\^\do\_\do\%\do\~}
\def\mn@doi{\begingroup\mn@urlcharsother \@ifnextchar [ {\mn@doi@}
  {\mn@doi@[]}}
\def\mn@doi@[#1]#2{\def\@tempa{#1}\ifx\@tempa\@empty \href
  {http://dx.doi.org/#2} {doi:#2}\else \href {http://dx.doi.org/#2} {#1}\fi
  \endgroup}
\def\mn@eprint#1#2{\mn@eprint@#1:#2::\@nil}
\def\mn@eprint@arXiv#1{\href {http://arxiv.org/abs/#1} {{\tt arXiv:#1}}}
\def\mn@eprint@dblp#1{\href {http://dblp.uni-trier.de/rec/bibtex/#1.xml}
  {dblp:#1}}
\def\mn@eprint@#1:#2:#3:#4\@nil{\def\@tempa {#1}\def\@tempb {#2}\def\@tempc
  {#3}\ifx \@tempc \@empty \let \@tempc \@tempb \let \@tempb \@tempa \fi \ifx
  \@tempb \@empty \def\@tempb {arXiv}\fi \@ifundefined
  {mn@eprint@\@tempb}{\@tempb:\@tempc}{\expandafter \expandafter \csname
  mn@eprint@\@tempb\endcsname \expandafter{\@tempc}}}

\bibitem[\protect\citeauthoryear{{Abbott} et~al.,}{{Abbott}
  et~al.}{2017}]{Abbott2017a}
{Abbott} B.~P.,  et~al., 2017, \mn@doi [\prl] {10.1103/PhysRevLett.119.161101},
  \href {https://ui.adsabs.harvard.edu/abs/2017PhRvL.119p1101A} {119, 161101}

\bibitem[\protect\citeauthoryear{{Acernese} et~al.,}{{Acernese}
  et~al.}{2015}]{Acernese2015}
{Acernese} F.,  et~al., 2015, \mn@doi [Classical and Quantum Gravity]
  {10.1088/0264-9381/32/2/024001}, \href
  {https://ui.adsabs.harvard.edu/abs/2015CQGra..32b4001A} {32, 024001}

\bibitem[\protect\citeauthoryear{{Ackley} et~al.,}{{Ackley}
  et~al.}{2020}]{Ackley2020}
{Ackley} K.,  et~al., 2020, arXiv e-prints, \href
  {https://ui.adsabs.harvard.edu/abs/2020arXiv200201950A} {p. arXiv:2002.01950}

\bibitem[\protect\citeauthoryear{{Alexander} et~al.,}{{Alexander}
  et~al.}{2017}]{Alexander2017}
{Alexander} K.~D.,  et~al., 2017, \mn@doi [\apjl] {10.3847/2041-8213/aa905d},
  \href {https://ui.adsabs.harvard.edu/abs/2017ApJ...848L..21A} {848, L21}

\bibitem[\protect\citeauthoryear{{Andreoni} et~al.,}{{Andreoni}
  et~al.}{2017}]{Andreoni2017}
{Andreoni} I.,  et~al., 2017, \mn@doi [\pasa] {10.1017/pasa.2017.65}, \href
  {https://ui.adsabs.harvard.edu/abs/2017PASA...34...69A} {34, e069}

\bibitem[\protect\citeauthoryear{{Antier} et~al.,}{{Antier}
  et~al.}{2020}]{Antier2020}
{Antier} S.,  et~al., 2020, \mn@doi [\mnras] {10.1093/mnras/staa1846}, \href
  {https://ui.adsabs.harvard.edu/abs/2020MNRAS.tmp.1969A} {}

\bibitem[\protect\citeauthoryear{{Arcavi} et~al.}{{Arcavi}
  et~al.}{2017}]{Arcavi2017}
{Arcavi} I.,  et~al., 2017, \mn@doi [\nat] {10.1038/nature24291}, \href
  {https://ui.adsabs.harvard.edu/abs/2017Natur.551...64A} {551, 64}

\bibitem[\protect\citeauthoryear{{Banerjee}, {Tanaka}, {Kawaguchi}, {Kato}  \&
  {Gaigalas}}{{Banerjee} et~al.}{2020}]{Banerjee2020}
{Banerjee} S.,  {Tanaka} M.,  {Kawaguchi} K.,  {Kato} D.,   {Gaigalas} G.,
  2020, arXiv e-prints, \href
  {https://ui.adsabs.harvard.edu/abs/2020arXiv200805495B} {p. arXiv:2008.05495}

\bibitem[\protect\citeauthoryear{{Barbieri}, {Salafia}, {Perego}, {Colpi}  \&
  {Ghirlanda}}{{Barbieri} et~al.}{2019}]{Barbieri2019}
{Barbieri} C.,  {Salafia} O.~S.,  {Perego} A.,  {Colpi} M.,   {Ghirlanda} G.,
  2019, \mn@doi [\aap] {10.1051/0004-6361/201935443}, \href
  {https://ui.adsabs.harvard.edu/abs/2019A&A...625A.152B} {625, A152}

\bibitem[\protect\citeauthoryear{{Bulla}}{{Bulla}}{2019}]{Bulla2019b}
{Bulla} M.,  2019, \mn@doi [\mnras] {10.1093/mnras/stz2495}, \href
  {https://ui.adsabs.harvard.edu/abs/2019MNRAS.489.5037B} {489, 5037}

\bibitem[\protect\citeauthoryear{{Bulla}, {Sim}  \& {Kromer}}{{Bulla}
  et~al.}{2015}]{Bulla2015}
{Bulla} M.,  {Sim} S.~A.,   {Kromer} M.,  2015, \mn@doi [\mnras]
  {10.1093/mnras/stv657}, \href
  {https://ui.adsabs.harvard.edu/abs/2015MNRAS.450..967B} {450, 967}

\bibitem[\protect\citeauthoryear{{Bulla} et~al.,}{{Bulla}
  et~al.}{2019}]{Bulla2019a}
{Bulla} M.,  et~al., 2019, \mn@doi [Nature Astronomy]
  {10.1038/s41550-018-0593-y}, \href
  {https://ui.adsabs.harvard.edu/abs/2019NatAs...3...99B} {3, 99}

\bibitem[\protect\citeauthoryear{{Chandrasekhar}}{{Chandrasekhar}}{1960}]{Chandrasekhar1960}
{Chandrasekhar} S.,  1960, {Radiative transfer}

\bibitem[\protect\citeauthoryear{{Christie}, {Lalakos}, {Tchekhovskoy},
  {Fern{\'a}ndez}, {Foucart}, {Quataert}  \& {Kasen}}{{Christie}
  et~al.}{2019}]{Christie2019}
{Christie} I.~M.,  {Lalakos} A.,  {Tchekhovskoy} A.,  {Fern{\'a}ndez} R.,
  {Foucart} F.,  {Quataert} E.,   {Kasen} D.,  2019, \mn@doi [\mnras]
  {10.1093/mnras/stz2552}, \href
  {https://ui.adsabs.harvard.edu/abs/2019MNRAS.490.4811C} {490, 4811}

\bibitem[\protect\citeauthoryear{{Coughlin} et~al.,}{{Coughlin}
  et~al.}{2020a}]{Coughlin2020a}
{Coughlin} M.~W.,  et~al., 2020a, \mn@doi [\mnras] {10.1093/mnras/stz3457},
  \href {https://ui.adsabs.harvard.edu/abs/2020MNRAS.492..863C} {492, 863}

\bibitem[\protect\citeauthoryear{{Coughlin} et~al.,}{{Coughlin}
  et~al.}{2020b}]{Coughlin2020b}
{Coughlin} M.~W.,  et~al., 2020b, \mn@doi [\mnras] {10.1093/mnras/staa1925},
  \href {https://ui.adsabs.harvard.edu/abs/2020MNRAS.497.1181C} {497, 1181}

\bibitem[\protect\citeauthoryear{{Coulter} et~al.,}{{Coulter}
  et~al.}{2017}]{Coulter2017}
{Coulter} D.~A.,  et~al., 2017, \mn@doi [Science] {10.1126/science.aap9811},
  \href {https://ui.adsabs.harvard.edu/abs/2017Sci...358.1556C} {358, 1556}

\bibitem[\protect\citeauthoryear{{Covino} et~al.,}{{Covino}
  et~al.}{2017}]{Covino2017}
{Covino} S.,  et~al., 2017, \mn@doi [Nature Astronomy]
  {10.1038/s41550-017-0285-z}, \href
  {https://ui.adsabs.harvard.edu/abs/2017NatAs...1..791C} {1, 791}

\bibitem[\protect\citeauthoryear{{Cowperthwaite} et~al.}{{Cowperthwaite}
  et~al.}{2017}]{Cowperthwaite2017}
{Cowperthwaite} P.~S.,  et~al., 2017, \mn@doi [\apj]
  {10.3847/2041-8213/aa8fc7}, \href
  {https://ui.adsabs.harvard.edu/abs/2017ApJ...848L..17C} {848, L17}

\bibitem[\protect\citeauthoryear{{Drout} et~al.}{{Drout}
  et~al.}{2017}]{Drout2017}
{Drout} M.~R.,  et~al., 2017, \mn@doi [Science] {10.1126/science.aaq0049},
  \href {https://ui.adsabs.harvard.edu/abs/2017Sci...358.1570D} {358, 1570}

\bibitem[\protect\citeauthoryear{{Evans} et~al.}{{Evans}
  et~al.}{2017}]{Evans2017}
{Evans} P.~A.,  et~al., 2017, \mn@doi [Science] {10.1126/science.aap9580},
  \href {https://ui.adsabs.harvard.edu/abs/2017Sci...358.1565E} {358, 1565}

\bibitem[\protect\citeauthoryear{{Fern{\'a}ndez}, {Tchekhovskoy}, {Quataert},
  {Foucart}  \& {Kasen}}{{Fern{\'a}ndez} et~al.}{2019}]{Fernandez2019}
{Fern{\'a}ndez} R.,  {Tchekhovskoy} A.,  {Quataert} E.,  {Foucart} F.,
  {Kasen} D.,  2019, \mn@doi [\mnras] {10.1093/mnras/sty2932}, \href
  {https://ui.adsabs.harvard.edu/abs/2019MNRAS.482.3373F} {482, 3373}

\bibitem[\protect\citeauthoryear{{Fern{\'a}ndez}, {Foucart}  \&
  {Lippuner}}{{Fern{\'a}ndez} et~al.}{2020}]{Fernandez2020}
{Fern{\'a}ndez} R.,  {Foucart} F.,   {Lippuner} J.,  2020, \mn@doi [\mnras]
  {10.1093/mnras/staa2209}, \href
  {https://ui.adsabs.harvard.edu/abs/2020MNRAS.497.3221F} {497, 3221}

\bibitem[\protect\citeauthoryear{{Foucart} et~al.,}{{Foucart}
  et~al.}{2013}]{Foucart2013}
{Foucart} F.,  et~al., 2013, \mn@doi [\prd] {10.1103/PhysRevD.87.084006}, \href
  {https://ui.adsabs.harvard.edu/abs/2013PhRvD..87h4006F} {87, 084006}

\bibitem[\protect\citeauthoryear{{Fujibayashi}, {Shibata}, {Wanajo}, {Kiuchi},
  {Kyutoku}  \& {Sekiguchi}}{{Fujibayashi} et~al.}{2020}]{Fujibayashi2020}
{Fujibayashi} S.,  {Shibata} M.,  {Wanajo} S.,  {Kiuchi} K.,  {Kyutoku} K.,
  {Sekiguchi} Y.,  2020, \mn@doi [\prd] {10.1103/PhysRevD.101.083029}, \href
  {https://ui.adsabs.harvard.edu/abs/2020PhRvD.101h3029F} {101, 083029}

\bibitem[\protect\citeauthoryear{{Goldstein} et~al.}{{Goldstein}
  et~al.}{2017}]{Goldstein2017}
{Goldstein} A.,  et~al., 2017, \mn@doi [\apjl] {10.3847/2041-8213/aa8f41},
  \href {https://ui.adsabs.harvard.edu/abs/2017ApJ...848L..14G} {848, L14}

\bibitem[\protect\citeauthoryear{{Gomez} et~al.,}{{Gomez}
  et~al.}{2019}]{Gomez2019}
{Gomez} S.,  et~al., 2019, \mn@doi [\apjl] {10.3847/2041-8213/ab4ad5}, \href
  {https://ui.adsabs.harvard.edu/abs/2019ApJ...884L..55G} {884, L55}

\bibitem[\protect\citeauthoryear{{Gompertz} et~al.,}{{Gompertz}
  et~al.}{2020}]{Gompertz2020}
{Gompertz} B.~P.,  et~al., 2020, \mn@doi [\mnras] {10.1093/mnras/staa1845},
  \href {https://ui.adsabs.harvard.edu/abs/2020MNRAS.497..726G} {497, 726}

\bibitem[\protect\citeauthoryear{{Haggard}, {Nynka}, {Ruan}, {Kalogera},
  {Cenko}, {Evans}  \& {Kennea}}{{Haggard} et~al.}{2017}]{Haggard2017}
{Haggard} D.,  {Nynka} M.,  {Ruan} J.~J.,  {Kalogera} V.,  {Cenko} S.~B.,
  {Evans} P.,   {Kennea} J.~A.,  2017, \mn@doi [\apjl]
  {10.3847/2041-8213/aa8ede}, \href
  {https://ui.adsabs.harvard.edu/abs/2017ApJ...848L..25H} {848, L25}

\bibitem[\protect\citeauthoryear{{Hajela} et~al.,}{{Hajela}
  et~al.}{2019}]{Hajela2019}
{Hajela} A.,  et~al., 2019, \mn@doi [\apjl] {10.3847/2041-8213/ab5226}, \href
  {https://ui.adsabs.harvard.edu/abs/2019ApJ...886L..17H} {886, L17}

\bibitem[\protect\citeauthoryear{{Hallinan} et~al.,}{{Hallinan}
  et~al.}{2017}]{Hallinan2017}
{Hallinan} G.,  et~al., 2017, \mn@doi [Science] {10.1126/science.aap9855},
  \href {https://ui.adsabs.harvard.edu/abs/2017Sci...358.1579H} {358, 1579}

\bibitem[\protect\citeauthoryear{{Higgins} et~al.,}{{Higgins}
  et~al.}{2019}]{Higgins2019}
{Higgins} A.~B.,  et~al., 2019, \mn@doi [\mnras] {10.1093/mnras/sty3029}, \href
  {https://ui.adsabs.harvard.edu/abs/2019MNRAS.482.5023H} {482, 5023}

\bibitem[\protect\citeauthoryear{{Hotokezaka}, {Nakar}, {Gottlieb}, {Nissanke},
  {Masuda}, {Hallinan}, {Mooley}  \& {Deller}}{{Hotokezaka}
  et~al.}{2019}]{Hotokezaka2019}
{Hotokezaka} K.,  {Nakar} E.,  {Gottlieb} O.,  {Nissanke} S.,  {Masuda} K.,
  {Hallinan} G.,  {Mooley} K.~P.,   {Deller} A.~T.,  2019, \mn@doi [Nature
  Astronomy] {10.1038/s41550-019-0820-1}, \href
  {https://ui.adsabs.harvard.edu/abs/2019NatAs...3..940H} {3, 940}

\bibitem[\protect\citeauthoryear{{Jin}, {Li}, {Cano}, {Covino}, {Fan}  \&
  {Wei}}{{Jin} et~al.}{2015}]{Jin2015}
{Jin} Z.-P.,  {Li} X.,  {Cano} Z.,  {Covino} S.,  {Fan} Y.-Z.,   {Wei} D.-M.,
  2015, \mn@doi [\apjl] {10.1088/2041-8205/811/2/L22}, \href
  {https://ui.adsabs.harvard.edu/abs/2015ApJ...811L..22J} {811, L22}

\bibitem[\protect\citeauthoryear{{Just}, {Bauswein}, {Ardevol Pulpillo},
  {Goriely}  \& {Janka}}{{Just} et~al.}{2015}]{Just2015}
{Just} O.,  {Bauswein} A.,  {Ardevol Pulpillo} R.,  {Goriely} S.,   {Janka}
  H.~T.,  2015, \mn@doi [\mnras] {10.1093/mnras/stv009}, \href
  {https://ui.adsabs.harvard.edu/abs/2015MNRAS.448..541J} {448, 541}

\bibitem[\protect\citeauthoryear{{Kasliwal} et~al.,}{{Kasliwal}
  et~al.}{2017}]{Kasliwal2017}
{Kasliwal} M.~M.,  et~al., 2017, \mn@doi [Science] {10.1126/science.aap9455},
  \href {https://ui.adsabs.harvard.edu/abs/2017Sci...358.1559K} {358, 1559}

\bibitem[\protect\citeauthoryear{{Kasliwal} et~al.,}{{Kasliwal}
  et~al.}{2019}]{Kasliwal2019a}
{Kasliwal} M.~M.,  et~al., 2019, \mn@doi [\mnras] {10.1093/mnrasl/slz007},
  \href {https://ui.adsabs.harvard.edu/abs/2019MNRAS.tmpL..14K} {p.~L14}

\bibitem[\protect\citeauthoryear{{Kawaguchi}, {Shibata}  \&
  {Tanaka}}{{Kawaguchi} et~al.}{2020}]{Kawaguchi2020}
{Kawaguchi} K.,  {Shibata} M.,   {Tanaka} M.,  2020, \mn@doi [\apj]
  {10.3847/1538-4357/ab61f6}, \href
  {https://ui.adsabs.harvard.edu/abs/2020ApJ...889..171K} {889, 171}

\bibitem[\protect\citeauthoryear{{Kiuchi}, {Sekiguchi}, {Kyutoku}, {Shibata},
  {Taniguchi}  \& {Wada}}{{Kiuchi} et~al.}{2015}]{Kiuchi2015}
{Kiuchi} K.,  {Sekiguchi} Y.,  {Kyutoku} K.,  {Shibata} M.,  {Taniguchi} K.,
  {Wada} T.,  2015, \mn@doi [\prd] {10.1103/PhysRevD.92.064034}, \href
  {https://ui.adsabs.harvard.edu/abs/2015PhRvD..92f4034K} {92, 064034}

\bibitem[\protect\citeauthoryear{{Korobkin}, {Rosswog}, {Arcones}  \&
  {Winteler}}{{Korobkin} et~al.}{2012}]{Korobkin2012}
{Korobkin} O.,  {Rosswog} S.,  {Arcones} A.,   {Winteler} C.,  2012, \mn@doi
  [\mnras] {10.1111/j.1365-2966.2012.21859.x}, \href
  {https://ui.adsabs.harvard.edu/abs/2012MNRAS.426.1940K} {426, 1940}

\bibitem[\protect\citeauthoryear{{Kyutoku}, {Ioka}  \& {Shibata}}{{Kyutoku}
  et~al.}{2013}]{Kyutoku2013}
{Kyutoku} K.,  {Ioka} K.,   {Shibata} M.,  2013, \mn@doi [\prd]
  {10.1103/PhysRevD.88.041503}, \href
  {https://ui.adsabs.harvard.edu/abs/2013PhRvD..88d1503K} {88, 041503}

\bibitem[\protect\citeauthoryear{{Kyutoku}, {Ioka}, {Okawa}, {Shibata}  \&
  {Taniguchi}}{{Kyutoku} et~al.}{2015}]{Kyutoku2015}
{Kyutoku} K.,  {Ioka} K.,  {Okawa} H.,  {Shibata} M.,   {Taniguchi} K.,  2015,
  \mn@doi [\prd] {10.1103/PhysRevD.92.044028}, \href
  {https://ui.adsabs.harvard.edu/abs/2015PhRvD..92d4028K} {92, 044028}

\bibitem[\protect\citeauthoryear{{LIGO Scientific Collaboration} et~al.,}{{LIGO
  Scientific Collaboration} et~al.}{2015}]{Ligo2015}
{LIGO Scientific Collaboration} et~al., 2015, \mn@doi [Classical and Quantum
  Gravity] {10.1088/0264-9381/32/7/074001}, \href
  {https://ui.adsabs.harvard.edu/abs/2015CQGra..32g4001L} {32, 074001}

\bibitem[\protect\citeauthoryear{{Lackey}, {Nayyar}  \& {Owen}}{{Lackey}
  et~al.}{2006}]{Lackey2006}
{Lackey} B.~D.,  {Nayyar} M.,   {Owen} B.~J.,  2006, \mn@doi [\prd]
  {10.1103/PhysRevD.73.024021}, \href
  {https://ui.adsabs.harvard.edu/abs/2006PhRvD..73b4021L} {73, 024021}

\bibitem[\protect\citeauthoryear{{Lamb} et~al.,}{{Lamb}
  et~al.}{2019}]{Lamb2019}
{Lamb} G.~P.,  et~al., 2019, \mn@doi [\apjl] {10.3847/2041-8213/aaf96b}, \href
  {https://ui.adsabs.harvard.edu/abs/2019ApJ...870L..15L} {870, L15}

\bibitem[\protect\citeauthoryear{{Li} \& {Paczy{\'n}ski}}{{Li} \&
  {Paczy{\'n}ski}}{1998}]{Li1998}
{Li} L.-X.,  {Paczy{\'n}ski} B.,  1998, \mn@doi [\apjl] {10.1086/311680}, \href
  {https://ui.adsabs.harvard.edu/abs/1998ApJ...507L..59L} {507, L59}

\bibitem[\protect\citeauthoryear{{Li} \& {Shen}}{{Li} \&
  {Shen}}{2019}]{Li2019b}
{Li} Y.,  {Shen} R.-F.,  2019, \mn@doi [\apj] {10.3847/1538-4357/ab2387}, \href
  {https://ui.adsabs.harvard.edu/abs/2019ApJ...879...31L} {879, 31}

\bibitem[\protect\citeauthoryear{{Lundquist} et~al.,}{{Lundquist}
  et~al.}{2019}]{Lundquist2019}
{Lundquist} M.~J.,  et~al., 2019, \mn@doi [\apjl] {10.3847/2041-8213/ab32f2},
  \href {https://ui.adsabs.harvard.edu/abs/2019ApJ...881L..26L} {881, L26}

\bibitem[\protect\citeauthoryear{{Manchado} et~al.,}{{Manchado}
  et~al.}{2004}]{Manchado}
{Manchado} A.,  et~al., 2004, in {Moorwood} A. F.~M.,  {Iye} M.,  eds,  Society
  of Photo-Optical Instrumentation Engineers (SPIE) Conference Series Vol.
  5492, Ground-based Instrumentation for Astronomy. pp 1094--1104,
  \mn@doi{10.1117/12.549188}

\bibitem[\protect\citeauthoryear{{Margutti} et~al.,}{{Margutti}
  et~al.}{2017}]{Margutti2017}
{Margutti} R.,  et~al., 2017, \mn@doi [\apjl] {10.3847/2041-8213/aa9057}, \href
  {https://ui.adsabs.harvard.edu/abs/2017ApJ...848L..20M} {848, L20}

\bibitem[\protect\citeauthoryear{{Matsumoto}}{{Matsumoto}}{2018}]{Matsumoto2018}
{Matsumoto} T.,  2018, \mn@doi [\mnras] {10.1093/mnras/sty2317}, \href
  {https://ui.adsabs.harvard.edu/abs/2018MNRAS.481.1008M} {481, 1008}

\bibitem[\protect\citeauthoryear{{Metzger}}{{Metzger}}{2019}]{Metzger2019}
{Metzger} B.~D.,  2019, \mn@doi [Living Reviews in Relativity]
  {10.1007/s41114-019-0024-0}, \href
  {https://ui.adsabs.harvard.edu/abs/2019LRR....23....1M} {23, 1}

\bibitem[\protect\citeauthoryear{{Miller} et~al.,}{{Miller}
  et~al.}{2019}]{Miller2019}
{Miller} J.~M.,  et~al., 2019, \mn@doi [\prd] {10.1103/PhysRevD.100.023008},
  \href {https://ui.adsabs.harvard.edu/abs/2019PhRvD.100b3008M} {100, 023008}

\bibitem[\protect\citeauthoryear{{Nakar}}{{Nakar}}{2019}]{Nakar2019}
{Nakar} E.,  2019, arXiv e-prints, \href
  {https://ui.adsabs.harvard.edu/abs/2019arXiv191205659N} {p. arXiv:1912.05659}

\bibitem[\protect\citeauthoryear{{Patat} et~al.,}{{Patat}
  et~al.}{2015}]{Patat2015}
{Patat} F.,  et~al., 2015, \mn@doi [\aap] {10.1051/0004-6361/201424507}, \href
  {https://ui.adsabs.harvard.edu/abs/2015A&A...577A..53P} {577, A53}

\bibitem[\protect\citeauthoryear{{Pian} et~al.}{{Pian} et~al.}{2017}]{Pian2017}
{Pian} E.,  et~al., 2017, \mn@doi [\nat] {10.1038/nature24298}, \href
  {https://ui.adsabs.harvard.edu/abs/2017Natur.551...67P} {551, 67}

\bibitem[\protect\citeauthoryear{{Rosswog}}{{Rosswog}}{2005}]{Rosswog2005}
{Rosswog} S.,  2005, \mn@doi [\apj] {10.1086/497062}, \href
  {https://ui.adsabs.harvard.edu/abs/2005ApJ...634.1202R} {634, 1202}

\bibitem[\protect\citeauthoryear{{Rosswog}, {Korobkin}, {Arcones}, {Thielemann}
   \& {Piran}}{{Rosswog} et~al.}{2014}]{Rosswog2014}
{Rosswog} S.,  {Korobkin} O.,  {Arcones} A.,  {Thielemann} F.~K.,   {Piran} T.,
   2014, \mn@doi [\mnras] {10.1093/mnras/stt2502}, \href
  {https://ui.adsabs.harvard.edu/abs/2014MNRAS.439..744R} {439, 744}

\bibitem[\protect\citeauthoryear{{Savchenko} et~al.,}{{Savchenko}
  et~al.}{2017}]{Savchenko17}
{Savchenko} V.,  et~al., 2017, \mn@doi [ApJL] {10.3847/2041-8213/aa8f94}, \href
  {https://ui.adsabs.harvard.edu/abs/2017ApJ...848L..15S} {848, L15}

\bibitem[\protect\citeauthoryear{{Siegel} \& {Metzger}}{{Siegel} \&
  {Metzger}}{2018}]{Siegel2018}
{Siegel} D.~M.,  {Metzger} B.~D.,  2018, \mn@doi [\apj]
  {10.3847/1538-4357/aabaec}, \href
  {https://ui.adsabs.harvard.edu/abs/2018ApJ...858...52S} {858, 52}

\bibitem[\protect\citeauthoryear{{Smartt} et~al.}{{Smartt}
  et~al.}{2017}]{Smartt2017}
{Smartt} S.~J.,  et~al., 2017, \mn@doi [\nat] {10.1038/nature24303}, \href
  {https://ui.adsabs.harvard.edu/abs/2017Natur.551...75S} {551, 75}

\bibitem[\protect\citeauthoryear{{Soares-Santos} et~al.,}{{Soares-Santos}
  et~al.}{2017}]{SoaresSantos2017}
{Soares-Santos} M.,  et~al., 2017, \mn@doi [\apjl] {10.3847/2041-8213/aa9059},
  \href {https://ui.adsabs.harvard.edu/abs/2017ApJ...848L..16S} {848, L16}

\bibitem[\protect\citeauthoryear{{Tanaka}, {Hotokezaka}, {Kyutoku}, {Wanajo},
  {Kiuchi}, {Sekiguchi}  \& {Shibata}}{{Tanaka} et~al.}{2014}]{Tanaka2014}
{Tanaka} M.,  {Hotokezaka} K.,  {Kyutoku} K.,  {Wanajo} S.,  {Kiuchi} K.,
  {Sekiguchi} Y.,   {Shibata} M.,  2014, \mn@doi [\apj]
  {10.1088/0004-637X/780/1/31}, \href
  {https://ui.adsabs.harvard.edu/abs/2014ApJ...780...31T} {780, 31}

\bibitem[\protect\citeauthoryear{{Tanaka} et~al.,}{{Tanaka}
  et~al.}{2018}]{Tanaka2018}
{Tanaka} M.,  et~al., 2018, \mn@doi [\apj] {10.3847/1538-4357/aaa0cb}, \href
  {https://ui.adsabs.harvard.edu/abs/2018ApJ...852..109T} {852, 109}

\bibitem[\protect\citeauthoryear{{Tanvir} et~al.,}{{Tanvir}
  et~al.}{2017}]{Tanvir2017}
{Tanvir} N.~R.,  et~al., 2017, \mn@doi [\apjl] {10.3847/2041-8213/aa90b6},
  \href {https://ui.adsabs.harvard.edu/abs/2017ApJ...848L..27T} {848, L27}

\bibitem[\protect\citeauthoryear{{Tinyanont}, {Millar-Blanchaer}, {Nilsson},
  {Mawet}, {Knutson}  et~al.}{{Tinyanont} et~al.}{2019}]{Tinyanont2019}
{Tinyanont} S.,  {Millar-Blanchaer} M.~A.,  {Nilsson} R.,  {Mawet} D.,
  {Knutson} H.,   et~al., 2019, \mn@doi [\pasp] {10.1088/1538-3873/aaef0f},
  \href {https://ui.adsabs.harvard.edu/abs/2019PASP..131b5001T} {131, 025001}

\bibitem[\protect\citeauthoryear{{Troja} et~al.,}{{Troja}
  et~al.}{2017}]{Troja2017}
{Troja} E.,  et~al., 2017, \mn@doi [\nat] {10.1038/nature24290}, \href
  {https://ui.adsabs.harvard.edu/abs/2017Natur.551...71T} {551, 71}

\bibitem[\protect\citeauthoryear{{Troja} et~al.,}{{Troja}
  et~al.}{2019}]{Troja2019}
{Troja} E.,  et~al., 2019, \mn@doi [\mnras] {10.1093/mnras/stz2248}, \href
  {https://ui.adsabs.harvard.edu/abs/2019MNRAS.489.1919T} {489, 1919}

\bibitem[\protect\citeauthoryear{{Utsumi} et~al.}{{Utsumi}
  et~al.}{2017}]{Utsumi2017}
{Utsumi} Y.,  et~al., 2017, \mn@doi [\pasj] {10.1093/pasj/psx118}, \href
  {https://ui.adsabs.harvard.edu/abs/2017PASJ...69..101U} {69, 101}

\bibitem[\protect\citeauthoryear{{Valenti} et~al.}{{Valenti}
  et~al.}{2017}]{Valenti2017}
{Valenti} S.,  et~al., 2017, \mn@doi [\apj] {10.3847/2041-8213/aa8edf}, \href
  {https://ui.adsabs.harvard.edu/abs/2017ApJ...848L..24V} {848, L24}

\bibitem[\protect\citeauthoryear{{Vieira} et~al.,}{{Vieira}
  et~al.}{2020}]{Vieira2020}
{Vieira} N.,  et~al., 2020, \mn@doi [\apj] {10.3847/1538-4357/ab917d}, \href
  {https://ui.adsabs.harvard.edu/abs/2020ApJ...895...96V} {895, 96}

\bibitem[\protect\citeauthoryear{{Watanabe} et~al.,}{{Watanabe}
  et~al.}{2018}]{Watanabe2018}
{Watanabe} M.,  et~al., 2018, in Ground-based and Airborne Instrumentation for
  Astronomy VII. p. 107023V, \mn@doi{10.1117/12.2311969}

\bibitem[\protect\citeauthoryear{{Watson} et~al.,}{{Watson}
  et~al.}{2019}]{Watson2019}
{Watson} D.,  et~al., 2019, \mn@doi [\nat] {10.1038/s41586-019-1676-3}, \href
  {https://ui.adsabs.harvard.edu/abs/2019Natur.574..497W} {574, 497}

\bibitem[\protect\citeauthoryear{{Watson} et~al.,}{{Watson}
  et~al.}{2020}]{Watson2020b}
{Watson} A.~M.,  et~al., 2020, \mn@doi [\mnras] {10.1093/mnras/staa161}, \href
  {https://ui.adsabs.harvard.edu/abs/2020MNRAS.492.5916W} {492, 5916}

\bibitem[\protect\citeauthoryear{{Wiersema} et~al.,}{{Wiersema}
  et~al.}{2012}]{Wiersema2012}
{Wiersema} K.,  et~al., 2012, \mn@doi [\mnras]
  {10.1111/j.1365-2966.2011.20379.x}, \href
  {https://ui.adsabs.harvard.edu/abs/2012MNRAS.421.1942W} {421, 1942}

\bibitem[\protect\citeauthoryear{{Yang} et~al.,}{{Yang}
  et~al.}{2015}]{Yang2015}
{Yang} B.,  et~al., 2015, \mn@doi [Nature Communications] {10.1038/ncomms8323},
  \href {https://ui.adsabs.harvard.edu/abs/2015NatCo...6.7323Y} {6, 7323}

\bibitem[\protect\citeauthoryear{{Zhu}, {Yang}, {Liu}, {Huang}, {Zhang}, {Li},
  {Yu}  \& {Gao}}{{Zhu} et~al.}{2020}]{Zhu2020}
{Zhu} J.-P.,  {Yang} Y.-P.,  {Liu} L.-D.,  {Huang} Y.,  {Zhang} B.,  {Li} Z.,
  {Yu} Y.-W.,   {Gao} H.,  2020, \mn@doi [\apj] {10.3847/1538-4357/ab93bf},
  \href {https://ui.adsabs.harvard.edu/abs/2020ApJ...897...20Z} {897, 20}

\makeatother
\end{thebibliography}

\end{document}